\RequirePackage[2018-12-01]{latexrelease}
\documentclass[final,review=false]{rQUF2e}

\usepackage{epstopdf}
\usepackage{subfigure}
\usepackage{color}
\usepackage{xcolor}

\theoremstyle{plain}

\usepackage[
backend=biber,
style=authoryear,
sorting=nyt,
]{biblatex}
\usepackage{pdflscape}
\usepackage{tabularx}
\theoremstyle{definition}

\theoremstyle{remark}

\newcommand{\highlight}{}
\newcommand{\highlightTwo}{}
\newcommand{\highlightt}{} 

\begin{document}


\title{Limit Order Book Simulations: A Review}

\author{KONARK JAIN$^{\ast}$$^\dag$$^\ddag$\thanks{$^\ast$Corresponding author.
Email: konark.jain.23@ucl.ac.uk \newline $\dag$ Opinions expressed in this paper are those of the authors, and do not necessarily reflect the view of JP Morgan.}, NICK FIROOZYE$^{\ddag}$, JONATHAN KOCHEMS$^{\dag}$ and PHILIP TRELEAVEN$^{\ddag}$\\
\affil{$\ddag$Department of Computer Science, University College London, London, UK\\
$\dag$Quantitative Research, JP Morgan Chase, London, UK} \received{v1.0 October 2023} }

\maketitle

\begin{abstract}
\highlight{Limit Order Books (LOBs) serve as a mechanism for buyers and sellers to interact with each other in the financial markets. Modelling and simulating LOBs is quite often necessary} for calibrating and fine-tuning the automated trading strategies developed in algorithmic trading research. The recent AI revolution and availability of faster and cheaper compute power has enabled the modelling and simulations to grow richer and even use modern AI techniques. In this review we \highlight{examine} the various kinds of LOB simulation models present in the current state of the art. We provide a classification of the models on the basis of their methodology and provide an aggregate view of the popular stylized facts used in the literature to test the models. We additionally provide a focused study of price impact's presence in the models since it is one of the more crucial phenomena to model in algorithmic trading. Finally, we conduct a comparative analysis of various qualities of fits of these models and how they perform when tested against empirical data.
\end{abstract}


The popularity of Limit Order Books in contemporary markets has been ever-rising. With real-time data access now provided by most exchanges and the rise of algorithmic trading systems, the order book and its history have become one of the most utilized forms of financial data. The reasons for this are plentiful; some of them include the complex nature of the dynamics of supply and demand, which are captured in the time evolution of the order book. The price formation of a security at the most granular level can be observed in the LOB, and the order book indicates the liquidity of the market, albeit not completely. Finally, the order book's history, in addition to the trade history, enables practitioners to effectively replay history and perform simulations and backtests. There have been several surveys focusing on the order book. The review on Limit Order Books by \cite{Gould2013} \highlight{focuses on studying the properties of the LOB but also showcases a number of } models for LOB simulation. In another survey \cite{cont2011} showed several zero-intelligence models' utility in LOB modelling and outlaid several empirical observations as tests for the model's outputs. \highlight{This survey focuses on the task of simulating the order book using historical data. We survey the recent developments across all major types of simulators and discuss each category's features and pitfalls. We focus on providing a brief summary of the methodology used in each simulation technique and provide a comparative study based on the stylized facts used to add priors to the simulator, test the simulator against empirical data, or both. We also perform an in-depth analysis of one particularly noteworthy aspect of LOB models - responsiveness to exogenous trades or Market/Price Impact.} 

\textbf{Motivation:} \highlight{There are a number of challenges in simulating the order book ranging from issues related to model complexity, difficulty in replicating the statistical properties of empirical data, and several mechanical issues stemming from the internal working of the exchanges such as halts in trading, open, intraday and close auctions, hidden orders, queue priority and dark pools. For an in-depth analysis of the challenges faced in LOB modelling, we refer the reader to \cite{Gould2013}. Despite, or due to, these challenges, modelling and simulating LOBs is of quite high importance for researchers and practitioners alike. An especially noteworthy use-case of a LOB simulator is for backtesting (or training) algorithmic trading strategies. The reason being having a simulator enables the availability of a richer data set for the strategy to run on and be refined on. Since each security's price has had just one realization of the various possible time evolutions of its LOB dynamics, if the trading strategy were to be fitted on just this one trajectory, issues of overfitting and thereafter lack of true out-of-sample performance will be apparent (\cite{white2000reality, sullivan1999data}). 


Avoiding overfitting ideally can be done by adding more data to the training set though generating more data is not trivial without knowing the generating process of the time series. One possible solution to this problem could be synthetic data using simulators. We note that training the strategies purely on a simulator might induce biases in the trading strategy since the simulator can never be perfectly representative of real data. This challenge can potentially be solved in two ways. The first and foremost is making sure the simulator is representative enough of the statistical properties of real world observed phenomena (i.e. `stylized facts'). This in itself encompasses the entire field of Order Book simulations - can a simulator be built which can replicate the distributions of the stylized facts observed in nature and at the same time is parsimonious? The second is the usage of real world data to do true out-of-sample testing of the trading strategy, or alternatively, combining the simulated data with real world data in the training data of the trading strategy. This will enable the practitioner to effectively ground their strategies in reality and avoid the pitfalls of backtesting on purely historical data as well as avoid inducing biases because of the simulator's lack of realism.}

\textbf{Contributions:} \highlight{Our contributions can be stated as follows. \begin{enumerate}
    \item We break down the types of models using their core modelling technique: Point Processes, Agent Based Modelling, Deep Learning, and models using Stochastic Differential Equations. In particular, there has been a recent rise of novel simulators with the onset of new generative modelling techniques such as Generative Adversarial Neworks  and its variants (\cite{goodfellow2014generative, mirza2014conditional, arjovsky2017wasserstein}).
    \item We study the variety of so-called `stylized facts', or empirically observed statistics of the LOB, that were used by the researchers as priors to develop their models and formulate a list of stylized facts which we believe are the more important ones for applications in algorithmic trading.
    \item We also highlight the various quality of fit tests done in each simulator and how they compare to each other. 
    \item Another important aspect we critique the literature is on the simulator's responsiveness to exogenous trades. This feature's importance stems from the fact that any practically applicable LOB simulator needs to be Market Impact aware as a zero Market Impact approximation may lead to poor out-of-sample performance (\highlight{\cite{biais1999price, foucault2005limit, cont2014price}}).
\end{enumerate} 

\section{Limit Order Books}

\highlight{In this section we provide a brief mathematical description on Limit Order Books and specifically how the time evolution of the order book dynamics can be described in a mathematical fashion. \highlightt{For a detailed overview, we refer the reader to \cite{abergel_anane_chakraborti_jedidi_munitoke_2016}.} The order book consists of discrete price levels at one `tick' difference from each other. This minimum difference in price levels is known as `tick-size' and is often specified by the regulators for each exchange. Each price level can have a non-negative integer number of `orders' resting there with different `sizes' and different `sides'. Orders here refer to the unexecuted (or 'unmatched') limit orders at that price level. Each order shows the intention of a market participant to trade at the specified price level, a quantity equalling the order size, and the direction of trade (buy or sell) specified by their order side. Since the order book consists of unmatched orders, the orders with intentions to buy are always at lower price levels than the order with intentions to sell. The buy section is know as `bid' and the sell section is known as `ask'. In Figure \ref{orderBookSnapshot}, for example, we show the aggregate order sizes at each price level for Apple at a random time of the day. 

The price level at best bid is known as `bid-price' : $P^B$ and similarly at best ask, we have `ask-price' : $P^A$. The distance in price units between bid and ask is known as `spread' : $S := P^A - P^B$. The `mid-price' is a theoretical price that signifies the average between $P^A$ and $P^B$: $P := \frac{P^A + P^B}{2}$. There are three categories of orders that can be placed in an order book: limit orders which are a buy (resp. sell) side order which has a price level equal or lower (resp. higher) than the lowest ask (resp. highest bid), market orders which are orders which match the requested bid/ask in the order book and remove the pre-existing limit order and finally cancel orders which are cancellations of limit orders without any execution. Note that in many markets, not all of the liquidity (i.e. unexecuted orders) is displayed in the LOB and we have the possibility of hidden orders and hidden executions or hidden trades. }

\begin{figure}[h]
\centering
\includegraphics[width=0.4\textwidth]{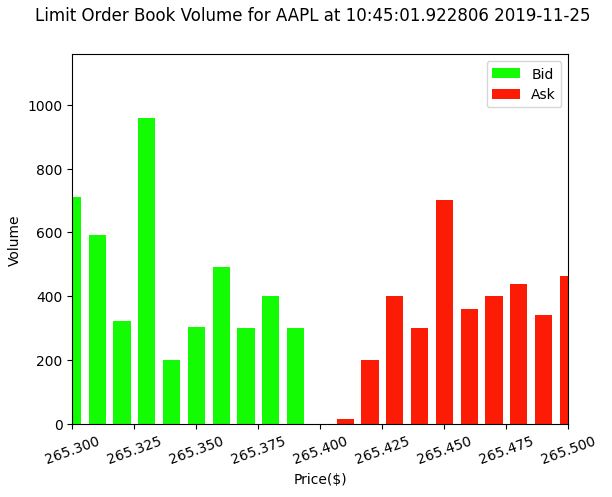}
\caption{Snapshot of a Limit Order Book for Apple on NASDAQ}
\label{orderBookSnapshot}
\end{figure}

\highlight{The order book is not stationary in time - it evolves with the three order types mentioned previously arriving randomly across the day. These orders evolve the order book which in turn evolves the price processes of the security. Hence, the LOB dynamics is the most granular level of price process formation. In Figure \ref{orderBookTimeEvolution}, we portray the top 10 levels of the LOB on each side and their time evolution for 5 minutes. The darkest hue corresponds to the best bid/ask and the lighter hues reflect the deeper levels. We also provide marks for trades (both visible and hidden). }

\begin{figure*}[h]
\centering
\includegraphics[width=0.8\textwidth]{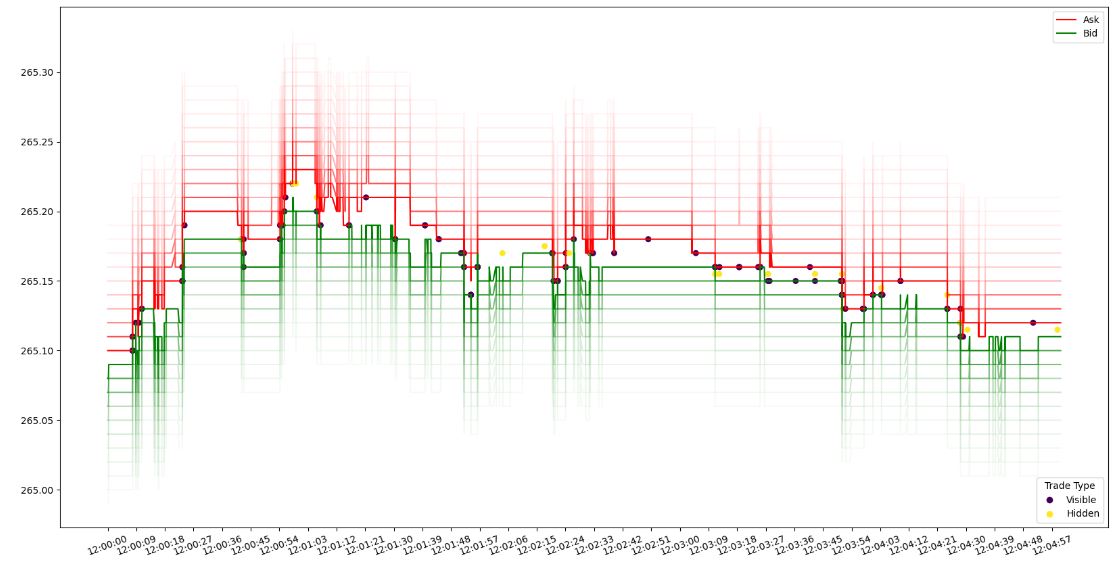}
\caption{Time evolution of the top 10 levels of a Limit Order Book for Apple on NASDAQ from 12:00 to 12:05 PM on 2019-11-25}
\label{orderBookTimeEvolution}
\end{figure*}

\highlight{\textbf{Dynamics:} We refer to, and adapt from, \cite{horst2017} in their treatment of the order book's time evolution by defining the order book state as: $$S_t (x) := (B_t, A_t, v_{b,t}(x), v_{a,t}(x))$$ where $B_t$ and $A_t$ are the best bid/ask prices respectively and $v_{b/a,t}(x)$ is the order book volume at $x$ (in units of the resp. currency) distance away from the mid-price $P_t$. $t$ here is the discrete time index. They define 8 types of events (bid/ask) which can change the state of an order book corresponding to Market Orders (bid/ask: \textbf{A/E}), Limit Order (in-spread) (bid/ask: \textbf{B/F}), Cancel Order \footnote{Horst and Paulsen assume Market Orders which do not deplete a price level are the same as Cancel Orders. Here we do not make that assumption.} (bid/ask: \textbf{C/G}) and Limit Orders (not in-spread) (bid/ask: \textbf{D/H}). The evolution of the order book at the event level can be described by the following: $$S_{t+1} = S_t + \mathcal{D}_t(S_t)$$ where $\mathcal{D}_t(.)$ is a random operator which depends on the dynamics that each 8 kinds of events induce on $S_t$. The dynamics induced are for example, if a queue-clearing buy/sell market order arrives, the ask/bid price moves up/down by one tick (or multiple ticks if the price levels near the best are empty), if a smaller market order arrives, it will change $v_{b/a, t}$ but not the prices, if an in-spread bid/ask limit order arrives, the bid/ask price moves up/down by one tick (or multiple ticks if the in-spread order is placed farther from best bid/ask) and so on for all 8 types of events. }

\highlight{The Order Book state is partially observable and is often modelled as a Markovian system. Several statistical properties of interest of the order books are studied in the literature and we provide a brief description on some of them relevant to order book modelling in the following section. }

\section{Stylized Facts}

\highlight{Stylized facts are the various observed statistical properties of the order book or one of the order book's features such as mid-price, spread, etc. Since the order book state itself is partially observable, studying the statistical properties of the order book dynamics is quite useful while designing simulators - both as priors for the model as well as empirical ground truth to test the goodness of fitness of the simulator. We enlist some of the most commonly used stylized facts below.}  

In their seminal paper \cite{Bouchaud2002} outline two of the most important stylized facts in literature - order flow statistics and average order book shape. \highlightTwo{\cite{cont2011}'s expansive survey lays down a number of other observations of the market:}

\begin{enumerate}

    \item \highlightTwo{Price changes, on a small enough timescale, are autocorrelated negatively at the first lag and then uncorrelated in further lags.}
    \item \highlightTwo{Trading volumes are heterogenous and strongly autocorrelated.}
    \item \highlightTwo{Trading volumes exhibit intraday seasonality strongly.}
    \item \highlightTwo{Order flow is clustered in time. This implies durations between orders are autocorrelated and there is positive cross-correlation among arrivals of different order types.}
\end{enumerate}

\highlight{These observations can be detected by measuring certain statistical properties of the LOB dynamics which are the so-called stylized facts. More recently, \cite{vyetrenko2020get} provide a list of recommended stylized facts for order book simulators. We briefly describe some stylized facts below: }

\highlight{
\begin{enumerate}
    \item \textbf{Empirical Distributions:}  densities of price, returns (particularly of note is the long-tailed distribution of returns), order volume, arrival rates, joint density of bid and ask queue sizes, time-to-fills.
    \item \textbf{Autocorrelation of returns:} $corr(r_{t + \tau, \Delta t}, r_{t, \Delta t})$ is the general formulation where $\Delta t$ is the time step of return calculation (which exhibits behaviours like vanishing autocorrelation of returns at larger time scales), alternatively the autocorrelation of absolute returns is another interesting aspect where we see a slower decay. Autocorrelation of the squared returns: $corr(r^2_{t + \tau, \Delta t}, r^2_{t, \Delta t})$ is generally taken as a measure to detect `Volatility Clustering'.
    \item \textbf{Correlations:} Volatility and Volume have positive correlations while Volatility and Returns have a negative correlation in empirical data.
    \item \textbf{Intraday seasonality:} Volumes have a signature U-shaped intraday seasonality with increased trading at Open and Close compared to mid-day.
    \item \textbf{Signature Plots:} This is defined as the relation between Volatility and Sampling Frequency - generally the empirical observation is that the signature plot decays quite slowly.
    \item \textbf{Average Shape of the Book :} The mean of volumes at each price level with respect to mid-price is the quantity of interest here. Generally we observed the so-called `M' shaped average shape for high spread stocks while a inverted `V' shape is observed for low spread stocks.
    \item \textbf{Price Paths:} The mid-price/ask-price/bid-price from the simulated order book is plotted against time in this stylized fact for a number of independent trials and comparisons are made to the empirical price paths observed. 
\end{enumerate}
}

\highlight{We further provide some details in Section 9 on the stylized facts used in the models we discuss in the following sections.}

\section{Point Processes Models}

\highlight{The order book in its essence can be thought of, mathematically, as an aggregate of several individual orders arriving at different point of times. It is hence quite natural to think of the order book as a queueing system and so there has been a plethora of models using Point processes to model the individual orders' arrivals. A point process has an associated counting process ($\equiv N_{t, t + \Delta t}$) which is the number of events occurring in $(t, t + \Delta t]$. With the usual conditions defined and satisfied on a complete probability space $(\Omega, \mathcal{F}, P)$, we define the intensity function $\lambda_t$ by: $$\lambda_t := \lim_{\Delta t \to 0} \frac{P(N_{t, t + \Delta t} > 0 | \mathcal{F}_t)}{\Delta t }$$ }

\subsection{Poisson Process and variants}

\highlight{Poisson Process modelling assumes that the order arrivals are independent of each other. There are several applications of the Poisson Process in the queueing systems literature so it is a very natural choice to model the LOB with a collection of Poisson Processes as well. Usually, the Poisson Process is classified as a `zero-intelligence model' i.e. a model which uses no prior information about the financial market or expert heuristics. However, as we will see later in this section, several ways of adding in priors have been formulated in the literature to make the Poisson Process models richer and more representative of the empirical observations.}

\textbf{\highlight{Zero-Intelligence Models:}} \cite{Bouchaud2002} show that using a zero-intelligence model, they are able to match the two stylized facts, \highlight{which are the shape of the order book and order flow arrival statistics,} they investigate quite well. \cite{Smith2003} treat the order flow of Limit Orders and Market Orders as sampling from a uniform probability distribution with Cancels occurring with a constant probability per unit time. They further develop a stochastic model of accumulated volumes from this order flow.

\cite{luckock2003} showed results of first order approximations' of the asymptotic behaviour of the Poisson arrival model. They show derivations of expressions for depth of the order book, time to fill and optimal order type. \highlightTwo{In \cite{ContStoikov2010}, the authors develop a model for the LOB by assuming Poisson arrival processes for Limit Orders, Market Orders and Cancellations. The arrival rates for Limit Orders \& Cancels depend inversely on the distance from opposite side's best quote, Cancels further depend on the number of outstanding shares at a level and Market Orders' arrival rate is considered to be constant. The authors show, using Laplace Transform, the probabilities for quantities of interest like direction of price move, making the spread and filling time conditional on the current state of the order book can be calculated analytically.} Building on \cite{ContStoikov2010}, \cite{gao2018hydrodynamic} use fluid approximations of the above model and form a law of large numbers for the order book shapes. Similarly \cite{kelly2018markov} also use fluid approximations to study a model with orders having Poisson arrivals with random price drawn from a stationary distribution for orders. \highlightTwo{Further in \cite{ContLerrard2013}, the authors show that with a Poisson arrival queueing system for quote dynamics, the price dynamics of the security can be thought of as a sum of independently and identically distributed (IID) random variables which, under the central limit theorem, forms a diffusion process.}  \highlightt{Finally, \cite{Abergel2011} show a detailed analysis of price dynamics converging to a Brownian motion with a Poisson queueing system model for the LOB. }

\textbf{\highlight{Variable order intensity Poisson models:}} \highlight{Moving away from the zero-intelligence approach, one way of adding priors to the Poisson model is to have non-constant order intensities.} \cite{hult2010algorithmic}, \highlight{for example}, use a Poisson arrival model with Limit and Cancel orders' intensities being dependent on distance from mid-price while Market Orders' intensity is kept constant. Further to sample the orders' sizes, they use a stationary exponential distribution. They use this model to building a Markov chain model of the LOB and further they create optimal trading strategies using it.

\cite{huang2015} study a variety of models for ask and bid queues around a fixed reference price - the first model they study is a \highlight{Poisson Process} model. They assume independence between ask and bid queues - here the arrival rates depend on  the current queue size instead of being constant. This is what they call a `queue-reactive' model \highlight{which is quite popular in practice}.   \highlightt{\cite{lu2018order} build on the above model and propose a non-Markovian order flow dynamic, albeit still using Poisson arrivals, by considering the order flow intensities to be dependent on not only the current state but also the previous history of order flow which led to this current state. They also address the limitations of having unit order size by consider limit orders sizes following a geometric distribution, cancels a truncated geometric and market orders a mixture of geometric distributions with Dirac delta functions for multiples of 50 to account for traders' preference over round numbers. They further proposed that in case of a queue depletion, the new limit order not only depends on the side of the cleared queue but also on the past removal events.  }

\textbf{\highlight{Discussion:}} \highlight{Despite their simplicity and vast variability,} Poisson arrivals do not fit well with some of the observed stylized facts. For example, the duration between orders are autocorrelated which leads to a clustering effect which Poisson processes are unable to explain. The core issue seems to be the assumption that all orders are independent which is generally contradictory to the practitioners judgement. Queue-reactive models do relax those assumptions to some extent but are still lacking in some other dimensions which we will highlight in the next section such as the endogeneity of the order arrivals. \highlightt{For a more detailed analysis, we refer the reader to  \cite{cont2011} and \cite{Abergel2011}. However due to their explainability and their mathematically convenient behaviour under scaling limits, Poisson models remain quite popular.} 

\subsection{Hawkes Process}
Hawkes Process as a way of modelling the LOB queueing system proves to be solving some of these challenges that Poisson Processes have. In their comprehensive review and tutorial, \cite{Bacry2015} outlay the major ideas of the Hawkes Process, its mathematical theory, some of its crucial properties and finally applications including a detailed review over the Order Book models. Furthermore, they provide insights into calibration methodologies for empirical fitting and testing. The major two areas of significant improvement we see in Hawkes Process methods compared to Poisson methods is first, volatility clustering effect is observed in Hawkes Models and second, the Epps Effect which is the zeroing down of covariance between two assets as well limit the timescale to zero. \highlight{We note that Hawkes Processes inherently have endogenously excited order flow as well as an implicit form of market impact. A more detailed study on the market impact of Hawkes process models is presented in Section 9.} More recently \cite{Hawkes2018} reviews the financial applications of Hawkes Processes. 

\textbf{Mathematical overview:} \highlight{Hawkes processes relax the assumption of independent increments in Poisson and instead use the fact that order flow is endogenously excited in its modelling. A multi-dimensional Hawkes process can also have cross excitation terms between the different dimensions (for eg, a 2D Hawkes Process of (ask volume, bid volume) can have 4 excitation terms - ask$\to$ask, bid$\to$bid, ask$\to$bid and bid$\to$ask). Let us briefly describe mathematically the Hawkes Process formulation here. The intensity function of the Hawkes process contains the self and mutual excitation terms mentioned previously. For a $d$-dimensional Hawkes process the intensity of the process $\lambda^{(i)}_t$ and the associated counting process $N^{(i)}_t$ for $ i = 1,\ldots, d$ is defined as: 
\begin{align}
    \lambda^{(i)}_t  = \mu^{(i)}_t + \sum^d_{j=1} \sum_{t_j \in T_j} \phi^{(j \rightarrow i)}(t - t_j)
\end{align} where $T_j := \{t_j : t_j \leq t\}$ denotes the set of past event times in the $j$ dimension of the Hawkes Process. Here, $\mu^{(i)}_t$ is the exogenous intensity of the $i$-th dimension and $\phi^{(j \rightarrow i)}(t - t_j)$ is the excitation term from $j$-th dimension to $i$-th dimension. The excitation terms are a function of the time since the event (generally a decaying function in time like exponential decay or power law decay). An alternate but similar formulation is the following: \begin{align}
    \lambda^{(i)}_t  = \mu^{(i)}_t + \sum^d_{j=1} \int^t_{0} \phi^{(j \rightarrow i)}(t - s) dN^{(j)}_s
\end{align} There have been several discussions in the literature on the choice of the kernel functions. \cite{nystrom2022hawkes} show that a power law kernel fits much better to the empirical data than the exponential kernels. \highlight{\cite{fonseca2014calib} also show using a Q-Q plot comparison between the distributions of empirical inter-arrival times to the exponential distribution that exponentially decaying kernels are probably insufficient in representing the empirical data.}

\textbf{n-dimensional Hawkes Process:} In recent years there has been a significant increase in LOB models using Hawkes Process albeit with vastly different formulations. \cite{toke2010market} creates a two-agent based model where liquidity takers (Market Orders), and liquidity providers' Limit Orders are each modelled as 1D Hawkes processes. Cancels are modelled to be Poisson arrivals and the price for Limit Orders and Cancels are sampled from a probability distribution. They compare performance improvement of using Hawkes processes against Poisson processes and show that Hawkes Processes with three excitations: Limit \& Market Orders' self-excitation and Market Orders exciting future Limit Orders give better fits to empirical data.  \cite{bacry2016estimation} divide the events in an order book into two categories - those which change the mid-price and those which do not. They use an 8-dimensional Hawkes process with orders changing the mid price being modelled by one dimension, and for events which don't change the mid price are modelled by 3 dimensions (Market Orders, Limit Orders and Cancellations). This is done for both bid and ask sides giving us a total of 8 dimensions. Previously, \cite{Large2007} followed a similar technique by using a 10D Hawkes Process: \{\{Limit Order, Market Order\} $\times$ \{change mid, don't change mid\}, \{Cancel Orders\}\} $\times$ \{bid, ask\}. They formalize the `resiliency' of the order book which is the ability of the order book to replenish after being depleted by a large trade. \highlight{\cite{kirchner2017estimation} proposed an alternative, non-parameteric way of estimating the Hawkes process and showed the applicability of their method in LOB data. Particularly noteworthy is the techniques they show in doing model-selection and their usage of the AIC statistic to optimize the hyperparameters. }

\highlight{\cite{fonseca2014calib} propose an alternate strategy to fit the Hawkes Process by using a generalized method of moments to fit the first four moments of various quantities of interest in the Hawkes Process. This method is claimed to be much faster than the traditional maximum likelihood estimation (MLE) methods used in the literature. They show weak convergence results of the fitted parameters to the true unknown parameters of the Hawkes Process. They use several key stylized facts to test the realism of their simulations. They also compare the fitted parameters to the MLE baseline.} 

\textbf{Constrained Hawkes Process:}  \highlightt{\cite{Zheng2014} create a 4 dimensional Hawkes Process for Level 1 Order Book simulation with two for each of ask and bid queues and they construct a spread process to control events where bid becomes greater than ask. They thus create and provide analysis for a Hawkes Process with constraints. }\cite{lee2022modeling} create a 4D Hawkes Process for level 1 LOB simulation similar to \cite{Zheng2014} but instead of constructing a separate spread process, they propose that the exogenous intensity for spread-narrowing events is a function of spread relative to the price. This implies that at 1 tick wide spread, the exogenous intensities of spread-narrowing events is zero. Further they also let the decay kernels' (for self and cross excitation) to be again dependent on spread but also stochastic. In this manner they ensure that the intensity of spread-narrowing events is exactly zero when the spread is 1 tick wide. They derive some properties of the price and spread process. Further, they provide some techniques of checking the estimator's bias and also provide a comprehensive empirical study and derive several economic explanations for the observed phenomena.

\textbf{Other variants:} \cite{kaj2017buffer} model the order book events in the following manner: Market Orders as queue-reactive Poisson arrivals, Limit Orders as Hawkes with excitation from Market Orders and Cancels as constant intensity Poisson arrivals. They call this model a `Buffer-Hawkes' Process. \cite{pakannen2022} develop a state-dependent Hawkes process where two types of states are considered: first, on the basis of spread being one tick or more than one tick, and second, on the basis order flow imbalance. They conclude that the excitation effects are observed to be highly dependent on the current state. They perform a number of analyses on their fitted Hawkes process to infer economic rationale behind observed effects in the kernels of the Hawkes process. \highlight{ \cite{kirchner2022hawkes} also formulate a marked state-dependent Hawkes Process but they use non-parameteric methods to estimate the excitation kernels shape (although they use power laws to fit the shape later). They use the current imbalance as the state indicator and also work towards creating a parsimonious model by zero-ing out the smaller excitations they observe in the data. } Another state-dependent model using Hawkes Process was proposed by \cite{mucciante2023estimation} where they consider the intensity as a product of a Hawkes Process driven intensity with a linear function on some observables in the market environment. We note that a key feature of this paper is the incorporation of time of day into its modelling - it is well known that order arrival intensities are non-stationary intraday and therefore most models clip the data to exclude open and close effects. \highlight{\cite{wu2019queue} take inspiration from \cite{huang2015} queue reactive model to build a Hawkes Process with exogenous intensity being queue-reactive in one model, and a queue reactive multiplier on top of the the Hawkes process intensity in the second model. They show that adding queue-reactiveness improves the goodness of fit of the models against Huang et al.'s model as well as the non-queue reactive Hawkes model. \cite{rambaldi2017role} show that the order size (and not just order count) of each individual order is important in the excitation of future orders. They show that using a marked Hawkes Process, with the marks corresponding to various bins of the order sizes, they are able to follow \cite{bacry2016estimation}'s non-parametric estimation technique to create a more realistic simulator.}

\textbf{Scaling limits:}  \highlightt{In their \highlight{influential} paper, \cite{Abergel2015} create a model of the full LOB where each level's Market Order and Limit Order intensities are modelled as Hawkes Processes with Cancels being modelled as having a queue-reactive Poisson intensity. They show through a mathematical analysis that this model can be used to create Stochastic Differential Equations (SDE) for aggregate features. } \cite{horst2019scaling} further showed that under some scaling limits, the Hawkes model for an LOB converges to an SDE for ask and bid prices while the intraday volume follows a system of Ordinary Differential Equations (ODE). They also show that the stationary intensities of the different types of events form Volterra-Fredholm Integral Equations. 

\textbf{Non-linear Hawkes Process:}  \highlightt{\cite{LuAbergel2018} create a 12D Hawkes process :  \{Limit Order, Market Order, Cancels\} $\times$ \{change mid, don't change mid\} $\times$ \{bid, ask\}. They compare the performance of a Linear and Non-Linear Hawkes Process with the non-linear one having novel \textit{inhibiting} kernels for negative excitation. They floor the intensities of the non-linear Hawkes model to zero. Another notable novelty in their research is the use of sum of exponential functions with varying half-lives as their kernels. }\cite{mounjid2019asymptotic} also create a non-linear Hawkes Process to simulate the order book with the non-linear transformation being dependent on the event type, the current state of the LOB, the current time and a sum over past events' excitations. These excitation kernels further are allowed to depend on the event type and the current state of the order book. They perform a mathematical analysis comparing this framework with different kinds of intensity models : Poisson, queue-reactive Poisson, Hawkes and Quadratic Hawkes processes.

\textbf{Neural Hawkes Process:} More recently, \cite{kumar2021deep} developed a Deep Neural Hawkes Process for Market Making in a simulated order book. The order book is simulated through a combination of agent based traders and sampling a Hawkes Process fitted to historical data. They use LSTMs to improve upon the Neural Hawkes Process proposed by \cite{mei2017neural}. Their hypothesis being that LSTMs are able to capture the more complex dynamics of feedback loops between various orders in the market since they inherently have this feature in their structure. A more detailed analysis of the various agents considered by the author is presented in Section 6. \highlight{ \cite{shi2022state} develop a neural Hawkes process with each order type's intensity being modelled by continuous time LSTM units. The process's intensity rates evolve in a way such that the current market state influences it. They draw the price and the size of the order from stationary distributions in their simulations.}

\textbf{Discussion:} \highlight{Hawkes process, with their high adaptability, provide a more comprehensive point process methodology to model the order book arrivals without having to necessarily model individual traders' behaviours in the market. Most importantly, their ability to reproduce important microstructural details like volatility clustering and Epps effect make them great candidates for the LOB models. It is of note that since these models (point processes) are mathematically descriptive, they are fully explainable in their nature and hence are suitable for applications where black-box solutions are not preferred. Recently, \cite{bacry2017tick} have published a Python library for calibrating Hawkes process. The key challenge that the practitioner might face in using Hawkes Process is the difficulty in the calibration of these models. This stems from the fact that the likelihood function is quite complex. In addition to that the choice of Kernels in the Hawkes Process can make quite a large difference in the model's predictive power. Further, the question of model parsimony becomes quite relevant here since the number of kernels scale in $O(n^2)$ for an $n-$dimensional Hawkes Process.}
\section{Agent based Models}

\highlight{Considering that the order book is constituted by a large number of heterogenuous agents, examples of heterogeneity being differences in their trade frequency, trading objectives, access to financial data and access to low latency trading hardware, the key idea in this category of LOB models is that each agent needs to be modelled in a separate category. For example, \cite{prenzelClient2023} make use of clustering techniques to showcase from anonymized trade execution data that there exists atleast four distinct clusters of types of agents in the lit LOB. The usual set of categories in addition to the usual informed v/s uninformed traders include high frequency traders, trend followers, mean reverters, noise traders and algorithmic traders.  \highlightt{We refer the reader to the expansive review by \cite{chakraborti2011econophysics} and Chapter 5 in the book by \cite{abergel_anane_chakraborti_jedidi_munitoke_2016} on Agent Based LOB models and review the newer research in the following.}}

\textbf{Recent Work:} \cite{Paddrik2012} use order speed and order placement as the differentiating characteristics to identify and model various types of agents. They use the May 6, 2010 \textit{Flash Crash} of E-Mini S\&P Futures to support their claim that many agents behave in a correlated manner. They create six categories of traders ranging for fundamental traders trading at very low frequency to market makers to HFTs. They model each category to be zero-intelligence Poisson processes. \cite{huang2015} in their second model assume that institutional agents post their limit orders at the top of the book while HFTs, market makers and arbitragers post it in deeper levels. Hence they propose that the order arrival intensity of a level depends on whether the level is the best bid/offer or not. Further they enhance this model by adding order arrival rate's dependency on opposite side queue size by discretising the opposite side queue size into 4 categorical quantiles. In their queue-reactive model they further relax assumptions by letting the mid-price or reference price change by one tick with some constant probability, and also having a reinitialization constant probability event. \cite{byrd2019abides} propose a software framework for simulating tens of thousands of agents with various types of objectives and trading patterns. They also introduce latency and an exchange agent for transactions to make the simulator more realistic. \cite{belcak2020fast} too create a software package like ABIDES and provide the a Python API with C++ backend of the simulations. They further study a number of simulation statistics and also provide a methodology to measure market impact with permanent and temporary components. 

\textbf{Combing ABMs with other methods: } \highlight{ \cite{lehalle2011high} describe the drawbacks of both the ABM (computational constraints and lack of analytical results) and Point Process Modelling (stationarity assumption and imperfect representation of stylized facts) approaches and propose using a mixed model. They create a zero-intelligence model (conditioned on a distance metric between the investor's view of the order book and the real order book) `pegged' to an ABM with scaling limits taken as a Mean Field Game. } As detailed in the previous section, \cite{kumar2021deep} uses a hybrid approach to modelling the LOB with Hawkes Process as their background process for several different types of agents to interact with. They propose to segregate the market participants into the following classes: the fundamental trader who follows a mean-reversion strategy, the chartist trader who follows a momentum strategy, a noise trader and three kinds of market-makers - one which uses their proposed Deep Hawkes Process to quote bid-ask orders, second which uses the Neural Hawkes Process proposed by \cite{mei2017neural}, and the third being a probabilistic market maker whose order placement is based on their view of the fundamental price of the security. \highlight{ \cite{shi2023neural} show that combing a stochastic model for a background simulator of the LOB with multi-agent simulation build on top of this background simulator has benefits over pure ABMs or pure stochastic models. They create a Neural Hawkes Process for the background simulator and use the ABIDES platform (\cite{byrd2019abides}) for the multi-agent simulation. They perform studies on price impact and observe herding behaviours in their simulations. } 

\textbf{Discussion:} \highlight{The interplay between the plethora of market participants has naturally led the LOB to be modelled in a statistical physics way. Agent Based Models of the LOB rise from the popularity of Econophysics modelling. The common difficulties with this kind of modelling is the heavy usage of heuristics in defining the behaviour of an individual or a class of agents. In addition to that, the computational cost of simulating individual agents is generally higher than other alternatives. Although, Mean Field Games analysis of the ABM system does help in the analytical tractability of this category of models. The usage of ABMs in combination with a background model is a promising area of future research since that combines the best out of both these constrasting modelling techniques. }

\section{Deep Learning based Models}

\highlight{Owing to the several sources of possible complexities and non-linearities in the time evolution of the order book as well as the distribution of the prices/returns and volumes, large parameterised models like deep learning networks have found recent surge in popularity in LOB simulation. There has been a significant amount of research done to use the predictive power of neural networks for predicting the mid-price, the volatility and the direction of price moves. Some of the more popular architectures being used are Convolutional Neural Networks (CNNs), Long Short-Term Memory networks (LSTMs), Recurrent Neural Networks (RNNs), and Generative Adversarial Networks (GANs). For a detailed description of these architectures, we refer the reader to the text by \cite{Goodfellow-et-al-2016}. For a focused review of machine learning applications, encompassing both the so-called traditional machine learning models and deep learning models, in finance, we refer the reader to the text by \cite{capponi_lehalle_2023}. }

\textbf{Mid-price prediction from LOB:} \highlightTwo{\cite{sirignano2018} use deep learning techniques like LSTMs to model the price formation mechanism with historical price and order flow as inputs. They show that their price dynamics is highly path-dependent since increasing performance was observed with increased history. Although they do not model the limit order book state explicitly but rather model the next mid-price which is just a property of the order book as a whole, their universality results show the promise of deep learning in ingesting tick data.}  \cite{Zhang_2019} use Deep Learning structures like CNNs coupled with LSTM and Inception modules to predict future price movements from the current state of the order book. \highlight{More recently, \cite{zhang2021deep} use Deep Learning on Market by Order data (Level 3 data) to predict the future price movements' category among up, down or flat.} A detailed comparative analysis on price prediction from LOB states is presented by \cite{briola2020deep}.

\textbf{Recurrent Neural Networks:} \highlight{ \cite{shi2021lob} make use of Recurrent Neural Network (RNN) structures like the Gated Recurrent Unit (GRU) and an ODE-RNN (Ordinary Differential Equation-RNN) to predict the volume at different levels of the order book. They use top of the book data (Level 1) to simulate five levels of data (Level 2). The authors claim that the ODE-RNN usage here is of particular importance since the traditional RNNs are unable to handle non-uniform time intervals in their history. Further they use transfer learning to show that the parameters learnt by training the network with one security's data can be fine tuned to a different security's data to get reasonably good performance. Further in \cite{shi2021limit}, they propose the usage of exponential decay kernels instead of the ODE kernels to make the model more parsimonious and to reduce the computational cost. They enrich their testing universe by using a wider set of stocks and they remove look-ahead biases from their previous model. They find that the order volume prediction accuracy decreases with increase in volatility.}

As mentioned in Sections 5 and 6, \cite{kumar2021deep} used LSTMs in their Hawkes Process model to capture more complex feedback loops dynamics which exist in the various event types in the market. 

\textbf{Generative Networks:} \cite{Takahashi2019}, \cite{Wiese_2020} and \cite{Ni2022} use GANs and its variants to generate financial time series. Particularly noteworthy is \cite{Wiese_2020}'s  use of the DY Metric (\cite{dymetric}) to test the performance of the generative network. \cite{wang2020} use conditional Wasserstein GANs to create their Stock-GAN model which simulates orders in the market by conditioning on some finite window of historical orders. They use an LSTM to encode the history and claim that the time dependence of order flow intensity is captured by this recurrent network. They further add a continuous double-auction approximation neural network to evolve the order book from the order streams that are simulated. \cite{lim2021intra} use Sequence GANs (SeqGANs) to model the order flow. They argue that the SeqGAN architechture is a better choice in handling discrete sequences of events like order flow than conventional GANs. They model the order book data using SeqGAN and further apply this simulation to do an analysis of the macro-level mid-price movements in this model. \highlightTwo{\cite{prenzel2022dynamic} created a methodology to calibrate GANs for order flow data dynamically for different market conditions instead of a single calibration over the entire dataset. They assume the order flow is following Poisson arrivals but rather than setting the intensity to a constant value, they use GANs to estimate the probability distribution function of the intensities for different conditions such as time of day and market volatility. \cite{cont2023limit} model the transitions between two consecutive LOB snapshots using Conditional Wasserstein GANs (conditional on current state of the LOB, therefore making the simulation Markovian). They particularly focus on creating a model which has implicit market impact in its order book transitions (more details in Section 9). They provide a thorough analysis of their model by comparing the simulation's and real world's empirical facts (more details in Section 10).}

\highlight{\cite{coletta2022learning} make use of Conditional GANs (CGANs) to create an LOB model (`world model') which is compared against a baseline ABM. They train the CGAN to generate the next trading action given the current features of the state of the order book. They further perform `Adversarial Attacks' on the CGAN model in \cite{coletta2023conditional} to highlight the dependence of the model on its input features. }

\textbf{Large Language Models:} \cite{nagy2023generative} use the recently popular autoregressive generative models on order book message data to simulate order flow. They tokenize the LOB messages and treat sequences of these tokens to simulate order flow as a Large Language Model (LLM) would treat words in a language to create a comprehensible sentence. They perform several out-of-sample tests on their simulator to test its efficacy.

\textbf{Discussion:} Deep Learning models for simulating the order book are natural candidates to solve for the vast complexity of the order book dynamics. The ability of deep neural networks to model convoluted time evolution of Markov processes coupled with the astronomical increase in recent years in ease of training such models has popularized this category of models. We note that a number of these models are able to reproduce the stylized facts quite well in their simulations. Some of them also exhibit the concave Market Impact characteristic that practitioners observe in the real world. However these models, like any other deep learning model, have many challenges such as the lack of explainability owing to their black-box nature, high sensitiveness to carefully calibrated hyperparameters and a very high model complexity with millions (or even billions) of parameters. A more parsimonious, explainable way of modelling the order book could be to model the transition of the order book state itself by taking the continuous time limit and creating a differential equation evolution of the stochastic process. 

\section{Stochastic Differential Equations Based Models}

\highlight{Since the LOB transitions are probabilistic in nature, the time evolution of an LOB state could be modelled as a set of differential equations. Here, usually, a continuous approximation is made of the state transition time which although is quite different from the reality of discrete time steps, but in larger time scales is an appropriate approximation since the frequency of LOB events is generally quite high. A major focus of this set of models is studying the long time dynamics of the order book - the so-called steady state (or the absence of it) is a major point of interest. Some of the more popular types of components used in the differential equations include diffusion and convection. These set of models are particularly important if one is concerned about the explainability of the LOB simulation. They also provide good segues into utilizing the modelled dynamics for optimal control problems like portfolio management, wealth management, optimal liquidation and market making. }

\textbf{Continuous limits of Point process models:} \cite{korolev2015modeling} model the order flow arrivals as a Cox process (doubly stochastic Poisson arrivals) and form stochastic differential equations for order flow imbalance. \cite{lakner2016} study one-sided order books assuming Poisson arrival of limit orders with intensity conditional on current best price. They provide weak limits of the price process and LOB process and show that the limit order book process is a solution to a stochastic differential equation. \cite{huang2017ergodicity} extend their previous Poisson arrival based framework \cite{huang2015} to create a more general stochastic dynamic model. They consider two separate jump processes for the order book state and a reference price which is not the mid-price but rather the so-called `fair-value' of the security as perceived by the traders. They create a state transition matrix based on a queue-reactive flow assumption and further also incorporate a reinitializaiton probability of the reference price attributing it to exogenous jumps. They perform tests of ergodicity and conclude with proving that under certain scaling limits, the price dynamics converges to that of a Brownian motion. More recently, \cite{cont2023mathematical} form a more general LOB model with the order flow modelled as a point process and the trade execution is modelled as a deterministic mass transport operator. In certain scaling limits, they show that their framework generalizes a number of other LOB models and show that they are in fact special cases of their framework.

\textbf{Volume of orders as a stochastic process:} \highlightTwo{\cite{ContLarrard2012} use heavy-traffic limits to show that depending on the scaling behaviour of the order flow, LOBs can act as deterministic under the fluid limit and stochastic under the diffusion limit. They find that the diffusion limit case occurs much more often in the empirical data. They come to the conclusion that the diffusion limit's differential equations' solution can be approximated by a two-dimensional Brownian motion. They further derive analytical solutions to questions like price dynamics, duration between price moves and probability of a price move.} Building on the model in \cite{ContLerrard2013}, \cite{chavez2017one} allow for variable spread in the simulated dynamics as well as allow for in-spread orders. They further formulate diffusion limits of the price process. \highlightTwo{In their influential paper, \cite{cont2021stochastic} develop a continuous limit of volume at a time $t$ and price $p$ by using a volume density $v(t,p)$. They centre the volume density to $u_t(x) := v(t, S_t + x)$ where $S_t$ is the mid-price. They follow a data-driven approach to model order book dynamics - they categorize the cancellations into deletions and modificiations. Further, modifications are bifurcated into symmetric (i.e. cancel and place at a nearby level) and antisymmetric (cancel and replace near mid-price). They model the symmetric modifications as a diffusion process, and the antisymmetric as a convection process.}

\textbf{Probabilistic properties under scaling limits:} \cite{horst2017}, by assuming generalized time-dependent order arrival intensities, develop limit theorems for price and volume densities at bid and ask. They conclude that given some regularity conditions, the two processes converge to a coupled ODE-PDE system of equations until scaling limits. \cite{horst2017weak} further generalize the previous work to develop a weak law of large numbers by considering the order flow as Markovian dynamics which are state dependent. Specifically they conjecture that the type of order, and its size \& price, all are a function of the price and standing volume of the order book. In \cite{horst2018second}, the authors further show that the previous model in \cite{horst2017weak} can be used as first order approximation of liquidity in the order book to construct optimal liquidation trajectories. They now develop second order approximations to obtain confidence intervals around these trajectories. To that extent, they formulate two scaling limits accounting for the empirical fact that price change fluctuations are much slower than order arrival and cancellation fluctuations. More recently, \cite{horst2023second} create a 2nd order approximation with a single scaling instead of the previous two. They do so by assuming an unvarying Market order to Limit order ratio. They show that the price-volume process in this limit converges to a solution of a infinite-dimension SPDE. \cite{ma2014dynamic, ma2022equilibrium} in their papers show that a one-sided order book dynamics can be modelled as an SPDE and equilibrium characteristics can be calculated for them. They further take the limit of traders in the market to infinity and show that a Mean-Field Game for the trader can be constructed using this SPDE, and under some conditions be solved using viscosity solutions of the Hamilton-Jacobi-Bellman (HJB) equations. On the other hand, \cite{rojas2020order} develop a law of large numbers, a central limit theorem and large deviations for a stressed order book - in their case they look at liquidity fluctuations. 

\textbf{Connecting various timescales:} \cite{hambly2020limit} build a set of models to connect the dynamics at various timescales - microscopic, then mesoscopic and finally, macroscopic. In the microscopic model they assume a Poisson arrival model for all order types with two intensities based on the frequency of trading: a common intensity for all order types is used for high frequency while an intensity dependent on queue price, mid-price, and the number of price changes before the current order is used at lower frequencies. Further they add a diffusion dynamic of orders diffusing to nearby price levels. They show that as the order arrival rate goes to infinity and the volume size of each order goes to zero, which is a continuous limit they formulate by looking at very small timescales, they can form a Markov diffusion process which is described by a system of reflected Stochastic Partial Differential Equations (SPDEs). They look at the time of price changes of these SPDEs to create the mesoscopic model of SPDEs. Finally, they take the limit of tick sizes going to zero to create a macroscopic continuous price process SPDE from the above model.

\textbf{Discussion:} \highlight{SPDEs provide a mathematically tractable formulation of the time evolution of the order book. This feature of this category of models makes them attractive but at the same time also have brings out certain issues. More often than not, the SPDEs do not have an explicit solution. There are several approximations made in the literature to circumvent this problem like using viscosity solutions, considering the similarities to the heat equation in physics, and even some inspiration from energy models from statistical physics. Despite the difficulty in finding exact solutions to the system of equations, these models can be used to build a simulator since it is readily possible to evolve a stochastic variable with a set of SPDE for its dynamics. We recommend to the practitioner that they be careful of the core assumptions of the SPDE model while they're using it. Some challenges in using these models include high model complexity and requirement of a high amount of computing resources to perform the simulations. We note that the class of point process models can be scaled for large timescales to produce a set of SPDEs of the model. Another desirable property of this category of models is that since the dynamics are analytically tractable, much like Hawkes Process models, we can use optimal control theory to participate in the simulator's trading.}

\section{Responsiveness to trades : Market Impact}

\textbf{Introduction:} Market Impact or Price Impact is defined as the price movement due to one's own trading. Suppose a large market order is submitted by an agent on the buy side and it depletes a few levels of prices in the order book, the new best ask price will be a few ticks higher than the previous best ask. This suggests that the agent's trade `walked the order book' and moved the price in the opposite direction. If the same agent wishes to buy again, they will have to pay a higher price. Similarly one can think of posting of limit orders shows the market one's intentions to trade at that price which gives \textit{information} to the market which can react against the agent. Reducing Market Impact has been one of the pillar stones of all agency and electronic trading activities with years of research spent on building algorithms to reduce the impact of large orders. Common techniques to reduce market impact include batching of orders, following the market (i.e. targeting the Volume Weighted Average Price) and using alpha signals to place orders at a \textit{smarter} price. Market Impact is one of the phenomena in the markets which cannot be measured immediately (atleast not in any meaningful sense) - the agent will probably wait for some time for the market to \textit{settle down} or in other words, wait for the price to come to back to the previous levels. This makes Market Impact one of the harder aspects of the LOB to model. Not only does it depend on the supply and demand of the order book at that point of time but also the volatility of the security. \highlightTwo{The importance of market impact has been highlighted in the literature since a long time (\cite{biais1999price, foucault2005limit, cont2014price}). Indeed Market Impact is, at the minutest of scales, quite closely related to the process of Price Formation (\cite{capponi_lehalle_2023_lillo}). \cite{cont2014price} show that the price impact can be explained by order imbalance and with a scaling they show that the 'square-root law' heuristic that traders in practice have can be derived quite easily. }

\textbf{Zero-Intelligence Models:} \cite{Smith2003} posit that the instantaneous price impact function $\phi(\omega, t)$ is nothing but the inverse of the cumulative depth profile $N(p, t)$ of the order book where $\omega$ is the order size. They show that by using Taylor's expansion on $\omega(\delta p)$, i.e. order size needed to move the mid price by $\delta p$, they get an analytical price impact function. Their results show good matching with observed price impact when the Taylor series is expanded to two degrees. Further they assume order arrival intensities are dependent on the distance from mid of a price level. In addition to that, they add an additive noise from the rapid submissions and deletions of High-Frequency traders in the form of a Brownian motion dependent on the centred volume density. They show that if the mid-price is an Arithmetic Brownian Motion, the volume density converges to a SPDE with a moving boundary problem's solution. They generalize this methodology to formulate an SPDE of the centred volume density and create a 2-factor model by creating SPDEs for both bid and ask side queues. They show that the long-term order book shape can be explicitly be solved for and they show a first order approximation of the shape in their results.

\textbf{Poisson process:} \cite{huang2015} study the market impact of VWAP liquidation and exponential scheduling liquidation in their simulated model. They see concavity in their market impact observations against time and volume both. This shows that their model has intrinsic market impact and it matches some real-world behaviour of the markets.  \highlightt{The same characteristics are observed in the model by \cite{LuAbergel2018}. In general, except for certain queue-reactive and state-dependent variants, Poisson models do not have market impact as a feature since the core assumption in a Poisson model is that each order event count is independent in increments.}

\textbf{Hawkes process:} As pointed out by Lillo (\cite{capponi_lehalle_2023_lillo}), given a reference price $P_t$, the time evolution of the price is a deterministic function of the order flow point process. They show that Market Impact can be modelled by the Transient Impact Model (i.e. trading velocity impacts the price in a decaying function of time) if the order flow is considered to be exogenous of the price process. However they argue that empirical data shows some correlation of price movement with future order flow and hence the assumption that order flow is exogenous to price impact is probably not correct. Hawkes process (with price and order flow as its dimensions) relaxes this assumption by considering cross excitation of order flow from price movement and vice-versa.

Hawkes models are by definition reactive to the past events and hence the order intensities are influence by any past orders happening. This can be thought of like an implicit form of market impact in these models.  \cite{bacry2015market} study the so-called Hawkes Impact Model by considering a simple 2D Hawkes process of price. They conclude that while liquidating a meta-order (i.e. an order made of multiple child-orders which are individual market/limit orders), they observe a concave market impact followed by a convex relaxation of the price after the agent has stopped trading. This behaviour is very much in line with the expectation of traders. Also noteworthy is the recent work by \cite{lee2017marked} where they study the market impact using a Hawkes process in a more realistic sense by including the tick-size discretization of price levels. They further develop formulae for realized volatility with this and compare it to empirical realized volatility. 

\textbf{Agent Based Models:} A recent study on how different categories of agents have different kinds of price impact has been done by Giamouridis et al. (\cite{capponi_lehalle_2023_Giamouridis}). \cite{Paddrik2012} provide an interesting case - they do not explicitly study the market impact of an exogenous order but they perform a similar study to replicate the circumstances around the Flash Crash of 2010 by placing a large exogenous trade that \textit{moves} the market. However we note that this impact is due to the behaviour of the High Frequency Traders and Market Makers modelled in the ABM and their reaction to the changing liquidity of the order book.  \cite{byrd2019abides} also show that agent based modelling can be utilised to study and estimate market impact models. They perform a case study with one exogenous trader placing orders, with varying proportions of Market and Limit orders, and they observe the price evolution during and after the trading is done. They perform comparisons of price paths with and without this exogenous trader and show that the price has meaningfully changed with the trading activity. The market impact they observe in their model is concave and is a decreasing function of the proportion of market orders in the trading strategy.  \cite{shi2023neural} follow the same framework to test the market impact in their model. \cite{coletta2022learning} too follow ABIDES framework of studying Market Impact. Further in \cite{coletta2023conditional} they breakdown the impact into the impact of Market Orders and Limit Orders and they compare it to historical replay method of order book simulations. \cite{belcak2020fast} study their model's market impact by analysing the average spread, variance of the spread and variance of the best price as a function of time since a large market order happened in the past. They also report a concave shaped market impact curve.

\textbf{Stochastic PDEs based Models:} \cite{horst2018second} show that using their 2nd order approximation, two different forms of Market Impact are found under different scaling limits - they term them to be temporary and permanent forms of impact. 

\textbf{Deep Learning based Models:} \highlightTwo{In  \cite{cont2023limit}, MI study is done by analysing the price paths observed while executing a varying quantity of orders using three strategies: Market Order TWAP (time weighted average price is the target price), Limit Order TWAP and POV (percentage of volume i.e. the trader targets maintaining their traded volume to be a constant ratio of the market volume). There are clear trend lines observed in all three and comparisons are made to Poisson and Hawkes in the former two where it is shown that there is no clear trend in Poisson or Hawkes. }


We see a general rise in order book models being sensitive to the subject of price impact however there still is room for improvement. We once again stress the importance of being aware of market impact in the order book simulators especially if the aim of building the simulator is to perform backtests of algotrading strategies. The ABIDES framework from \cite{byrd2019abides} proves to be quite useful in the study of market impact since it is available in an open source code repository.   

\section{Comparative study}

As we have seen above, there exists a rich literature spanning a number of modelling methods to model and simulate the order book. More often than not, we see researchers investigating some of the stylized facts observed in empirical data and using their reasonings for the observed distributions as priors in their modelling technique. It makes sense that building from those priors, they use these stylized facts as goodness-of-fit metrics as well. Table \ref{table:1}\footnote{
Glossary:

\begin{enumerate}
    \item PDF :  Probability Distribution Function
    \item Exp(.) : Exponential Distribution
    \item DL : Deep Learning
    \item ABM : Agent Based Model
    \item  MI : Market Impact
    \item SPDE: Stochastic Partial Differential Equations
    \item ACF : Auto-Correlation Function
\end{enumerate}
} enumerates the stylized facts being used in the models we review here. We also provide certain comments on how the authors have utilized the stylized facts in testing their model's efficacy against real data. 

The vast majority of researchers make use of empirical probability distribution functions of various properties of the order book, the most popular ones being inter-order arrival times, spread and volumes, as their primary stylized fact. The technique for testing against this stylized fact is usually a qualitative test where the two distributions (empirical data and simulations) are plotted against each other. While this method is useful to test whether the general shape of the distribution is matching between the two datasets, we also note the usage of the far more sensitive to tail events method of Q-Q plots in several papers. This technique not only matches the dense regions of the distribution but also the lower and higher quantiles. Indeed, most of the Hawkes Process methodology works we have mentioned in Table \ref{table:1} make use of the Q-Q plot on the inter-order arrival times stylized fact to refute the Poisson method. Some notable observations are mentioned below:

 \cite{Farmer2005} test the model proposed by \cite{Smith2003} by using empirical data from the London Stock Exchange. Notably their zero-intelligence model is able to reproduce a concave Market Impact function. \highlightTwo{\cite{ContStoikov2010} test their model's quality of fit by comparing the average LOB profile and realized volatility against real-world data from the Tokyo Stock Exchange. They further show that the conditional probability estimates from their model matches empirical frequencies observed of direction of price moves and one-step transition.} \cite{Abergel2011} demonstrate a series of tests to compare simulation results to real-world data. These included comparisons of average depth profile, probability distributions of spread in ticks and price changes, autocorrelation of price changes and Q-Q plot of mid-price changes.

\cite{Paddrik2012} show that their agent based approach does show volatility clustering phenomenon which is quite remarkable. It seems like a mixed timescales approach to modelling the LOB naturally leads to volatility clustering. Interestingly they're also able to simulate crashes in the market when a large sell order is traded and HFTs and Market-makers withdraw from the market.  

\cite{huang2015}'s three models are tested against empirical data for two high spread-in-tick stocks in the French exchange. The zero-intelligence Model I fits better to the asymptotic order distribution of these stocks compared to the constant arrival rate models. Interestingly, in their third model with moving mid-price, without a reinitialisation event, they find that the realized volatility in simulations was much lower than the empirical realized volatility. They conclude that their model probably suffers from a mean-reverting behaviour to mid-price which is not necessarily true in practice.

\cite{wang2020} perform tests on distributions of price, quantity, inter-arrival time and spectral bid/ask prices. This is of note since instead of using the qualitative tests methodology we commented on previously, they make use of statistical tests to provide a more robust testing methodology. Another notable mention is \cite{kirchner2017estimation}'s method of model selection. They provide some hyperparameters in their model to tune for each usecase and show examples of using the AIC metric in doing model selection in choosing these hyperparameters. An impressive study on the calibrated results is performed by \cite{lee2022modeling} where they perform stationarity checks on their parameters and also compare the calibrated parameters against a baseline model. They also perform model-selection by comparing the five different proposed models. \cite{mucciante2023estimation} showcase a testing methodology in which they measure the speed of convergence in their fitting method as well as show how sensitive their parameters are to various conditions to check the robustness of their model. They also perform tests on out-of-sample data. Comparisons to baseline models are also quite beneficial when the practitioner wishes to justify the incremental performance gain against simpler, more explainable methods. A good example are the papers \cite{shi2021limit, shi2021lob} where the authors compare their method with a number of low complexity baselines. They also perform several studies to validate their use of ODE-RNNs instead of the more traditional LSTMs/GRUs and further provide an ablation study on the parameters.

 \highlightt{As can be seen from Table \ref{table:1} Poisson models, generally, are successful in representing a number of `first order' stylized facts like distribution of spread, volumes, average depth, average order book profile. However \cite{Abergel2011} show that several other key stylized facts such as Autocorrelation functions of price changes, signature plots and long term volatility are insufficiently replicated in a Poisson model. They show that a self-exciting process model like Hawkes Process could be one of the candidate solutions to these. }

Hawkes Process models are seen to be much better than Poisson in representing the above mentioned stylized facts. They also fit the tails of the distribution of inter-order arrival time quite well. It has been shown \cite{pakannen2022} that the residuals after fitting the Hawkes process should follow the Exponential distribution however we see in \cite{pakannen2022} that this is generally not the case with the generic Hawkes Process. Therefore more complex models such as state-dependent Hawkes, Neural Hawkes Process etc are proposed. A frequent property we see being tested across all Hawkes Process models is the nature of the excitation kernel. The most usual choice of Exponential Kernels have been shown to be insufficient \cite{nystrom2022hawkes} and hence Power Law Kernels have become more popular \cite{bacry2016estimation} although they are harder to calibrate. A separate class of Hawkes models is the non-parametric estimation Hawkes models where the excitation kernels are estimated without any prior shape assigned to it. A number of authors report the shape of these kernels in time (or log-time). 

In the Deep Learning category of models, we see the training and validation losses being generally reported in the papers which is the general industry standard. Also notable, particularly in the models using GANs, is the usage of simulated price paths and their qualitative comparisons to the empirical data. The Stochastic Partial Differential Equations category of models generally use the stylized facts to create priors in their differential equation dynamics and we see a general trend towards testing the first order features and long term asymptotic features.  

Among the many mentioned stylized facts below, we suggest the reader to choose or formulate the ones which are the most important for their use-case. It is generally quite useful to do exploratory data analysis on your dataset before performing any data study, and calculating the various stylized facts is a good way to do the same.  We have listed some of the most popular ones in Section 3. While these stylized facts are universal in their importance, the specific distribution of these stylized facts varies across securities, asset classes and regions. We also stress the importance of comparing the model's efficacy against both real world data and simpler baselines as well as the current state of the art in the relevant category. This would enable the user to further understand the reason their model performs better or worse than other models. Finally, robustness of the model's calibrated parameters should be checked against initial conditions, market volatility and other quantities which can bring in elements of non-stationarity in the probability space one is estimating.

\section{Conclusion \& Future Work}

The field of order book simulations is growing in step with the modelling techniques themselves. For instance, the recent rise in Deep Learning's popularity has meant a number of deep architectures being specifically build for mimicking the order book and its properties. In this review we present a classification of the order book simulator models by their core modelling methodology. The view that the order book is a stochastic process with several components and that intensities of these components can be modelled either on their own (zero-intelligence models) or with interacting terms (Hawkes process) is the basis of the Point Process class of models. This class of models look at an aggregate view of the order book dynamics and are not concerned with the motivation or purpose behind each order. The dynamics are purely dependent on the current state of the LOB and the history. The polar opposite view is taken by the Agent Based Modelling set of models. Here the research focuses on mimicking the behaviour of the various kinds of agents in the real world interacting with each other and with the central limit order book. The challenges here are of course the vast variability of the kinds of agents real markets have as well as dependence of the simulator's realism on the assumptions made about the agent's expected behaviour. A third kind of modelling technique is to abandon forming any priors whatsoever on the data and let a universal approximator such as a deep learning network perform the estimation. The two major kinds of deep learning architectures we see in LOB modelling are firstly, the recurrent neural networks which predict the next order book state conditional on the history of the order book by performing non-linear transformations on the so-called memory of the order book states to produce the next one, and secondly, the generative networks where the underlying probability distribution function or the state transition probability is directly estimated. This technique, although quite powerful, has been shown to have difficulties in training and have high dependence on the hyperparameters. Deep learning's success does come with a major drawback - the lack of model explainability. Finally, the most explainable class of models are the ones which, using priors from observed stylized facts or expert judgement, use partial differential equations to model the dynamics of the order book states. These are known as the stochastic partial differential equation models and though they possess mathematical tractability which gives the user the ability to perform asymptotic analyses and solve optimal control problems in the order book framework itself, these models generally are seen to be overdependent on their prior assumptions.  

We provide a concentrated study on some of the more important statistical properties of the order book in Section 3 and further provide a comparative analyses of how the models we review are making use of these properties in both their model building as well as testing in Section 9. Finally we provide an in-depth analysis of the phenomena of market impact and how the simulators we study are sensitive to exogenous trades. 

Despite having such a wide variety of simulators, there is a lack of a parsimonious, explainable, analytically tractable model which has a good representation of most of the stylized facts the model is aiming to track. In our future work as a research group, we plan to tackle this challenge and build a simulator which can be interacted with by an autonomous agent to learn trading strategies on. 

\section{Disclaimer}

Opinions and estimates constitute our judgement as of the date of this Material, are for informational purposes only and are subject to change without notice. This Material is not the product of J.P. Morgan’s Research Department and therefore, has not been prepared in accordance with legal requirements to promote the independence of research, including but not limited to, the prohibition on the dealing ahead of the dissemination of investment research. This Material is not intended as research, a recommendation, advice, offer or solicitation for the purchase or sale of any financial product or service, or to be used in any way for evaluating the merits of participating in any transaction. It is not a research report and is not intended as such. Past performance is not indicative of future results. Please consult your own advisors regarding legal, tax, accounting or any other aspects including suitability implications for your particular circumstances. J.P. Morgan disclaims any responsibility or liability whatsoever for the quality, accuracy or completeness of the information herein, and for any reliance on, or use of this material in any way.

Important disclosures at: www.jpmorgan.com/disclosures

\clearpage
\begin{landscape}
\tiny
\begin{table}[!]
\begin{tabularx}{\linewidth}{|>{\tiny \hsize=0.175\linewidth}X|
                              >{\tiny \hsize=0.075\linewidth}X|
                              >{\tiny \hsize=0.375\linewidth}X|
                              >{\tiny \hsize=0.375\linewidth}X|}
    \hline
     \textbf{Model} & \textbf{Category} & \textbf{Stylized Facts Tested } & \textbf{Quality of Fit Comments} \\
    \hline
    \cite{Bouchaud2002} & Poisson  & Order flow statistics and average order book shape & Qualitative tests performed \\
    \hline
     \cite{Farmer2005} (tests \cite{Smith2003}) & Zero-intelligence & Average Spread, Price Diffusion rate  & All stylized facts compared between empirical and predicted.  \\
     
    \hline
     \cite{luckock2003} & Poisson  & PDFs of trade andbest ask/bid prices; Density of unexecuted orders w.r.t. price  & Qualitative tests performed  \\
    \hline
     \highlightTwo{\cite{ContStoikov2010}} & Poisson  & Arrival rates w.r.t. distance from opposite quotes; Average LOB Shape; Probability of price increase w.r.t. Queue Size; Probability of execution (one side and both sides) before mid-price movement  & Tests done to compare empirical simulated results for probability of price increase w.r.t Queue Size and across various levels. Comparisons are made with the empirical data as well as theoretical results by the Laplace transform method. \\
    \hline
      \highlightt{\cite{Abergel2011}}  & Poisson & Average Depth Profile, PDF of Spread-in-ticks; ACF of price increments, Price Paths, Signature Plots, Average Depth, Spread, long term volatility & Qualitative Tests Performed on all stylized facts; Q-Q tests are performed on mid-price increments. \\
    \hline 
     \highlightTwo{\cite{ContLerrard2013}} & Poisson & Joint distribution of best bid and ask size, Probability of price increase conditional on current bid/ask sizes   & Diffusive coefficient of Price from simulation and from empirical data is compared  \\
    \hline
    \cite{huang2015} & Poisson  & Order intensities distribution by queue size; PDF of queue size for 3 levels of bid/ask; Joint distribution of first two levels queue sizes; Joint distribution of best bid and ask queue sizes& Qualitative tests performed  \\ 
    \hline
    \highlightt{ \cite{lu2018order}}& Poisson & Order intensities and Order size distribution by queue size; Conditional distributions and other statistics of various types of orders and their arrival times; PDF of best bid/ask sizes& Qualitative tests performed using Monte Carlo simulations \\ 
    \hline
    \cite{toke2010market}  & Hawkes & PDF of spread, inter-arrival times, variance of mid-price & Qualitative tests performed  \\
    \hline
    \cite{Zheng2014} & Hawkes & Signature Plots (Bid1, Ask1, Mid) & Qualitative tests performed  \\
    \hline
    \cite{fonseca2014calib} & Hawkes & PDF and histograms of inter-arrival time; Autocorrelation function of number of trades in a time window; Signature plots  & Q-Q plot comparison made of inter-arrival time to the exponential distribution. Qualitative tests upon the stylized facts is performed. Comparisons of the fitted parameters with an MLE baseline is done along with testing of robustness of the params by investigating the standard deviations.  \\
     \hline
     \cite{bacry2016estimation} & Hawkes &  Event cross and self-excitation versus time & N/A  \\
    \hline
    \cite{rambaldi2017role}& Hawkes & PDF of inter-arrival times, order volumes & They show a number of plots and calibration results for their estimated Hawkes Process  \\
    \hline
     \highlightt{\cite{LuAbergel2018}} & Hawkes & Conditional probabilities of events; Signature Plots (mid-price); PDF of inter-event times & Residuals' Q-Q Plot tested to follow Exp(1) distribution; Qualitative tests performed \\
    \hline
    \cite{mounjid2019asymptotic} & Hawkes & No Tests Performed & N/A  \\
     \hline
    \cite{wu2019queue} & Hawkes & PDF of queue size & Q-Q plot of inter-arrival times are compared with empirical data as well as a Queue-reactive Poisson model baseline. Calibrated results are showcased for the model. Qualitative tests performed over stylized facts. \\     
    \hline
    \cite{kumar2021deep} & Hawkes; DL & PDF and Auto-correlation function of returns & Qualitative tests performed  \\
    \hline
    \cite{kirchner2022hawkes} & Hawkes & Average order intensity by time of day, unconditional transition probabilities between all orders, Market Orders against Imbalance & Properties of calibrated excitation kernels is plotted against time of day \\
    \hline
    \cite{pakannen2022} & Hawkes &  State transition probabilities & Residuals' Q-Q Plot tested to follow Exp(1) distribution. Comparison made between generic Hawkes and state-dependent Hawkes models. \\
    \hline
    \cite{lee2022modeling} & Hawkes & Bid/Ask Price plots  & Stationarity of estimated parameters is checked, comparisons between normal Hawkes and spread conditioned Hawkes (proposed model) is also done. Tests are also performed in further model selection between the 5 proposed models. Residuals' Q-Q Plot tested to follow Exp(1) distribution. \\
    \hline
    \cite{nystrom2022hawkes} & Hawkes & Price change in ticks frequency; Price Paths; PDFs of number of jumps  & Price Paths are compared between exponential and power law kernels. Statistical significance tests as well Qualitative tests on estimated parameters are also done. Further the compurational time for both types of Kernels' fitting process is reported. Distributions of number of jumps are compared qualitatively between empirical data and simulated paths.  \\
    \hline

\end{tabularx}
\label{table:1}
\end{table}
\end{landscape}
\clearpage

\begin{landscape}
\tiny
\begin{table}[!]
\begin{tabularx}{\linewidth}{|>{\tiny \hsize=0.175\linewidth}X|
                              >{\tiny \hsize=0.075\linewidth}X|
                              >{\tiny \hsize=0.375\linewidth}X|
                              >{\tiny \hsize=0.375\linewidth}X|}
    \hline
     \textbf{Model} & \textbf{Category} & \textbf{Stylized Facts Tested } & \textbf{Quality of Fit Comments} \\
    
    \hline
    \cite{shi2022state} &  Hawkes; DL & Price Paths; Volatility Clustering; Empirical PDFs of inter-arrival times; Volume - Volatility and Log returns - Volatility correlations   & Tests are performed to compare the model against several simpler baseline models.  \\
     \hline
    \cite{mucciante2023estimation}& Hawkes & N/A  & Convergence speed and Sensitivity to parameters are tested. Out of sample testing is also done using statistical significance tests.  \\
     \hline
    \cite{Paddrik2012}& ABM &  Volume \% by Agent type; Cancelation rates by Agent type; PDF of returns; Autocorrelation Function of absolute returns (indicative of Volatility Clustering); Autocorrelation Function of returns; PDF of returns sampled at various frequencies &  Q-Q Plots of Returns v/s Gaussian Distribution tested \\
    \hline
    \cite{byrd2019abides} & ABM & Price Paths; Trade Paths; & Price Paths are compared qualitatively. Errors and accuracy Out of Sample are also reported.  \\
    \hline
    \cite{belcak2020fast} & ABM & N/A & Performance comparisons are made with the ABIDES \cite{byrd2019abides} platform  \\
    \hline
    \cite{shi2023neural} & ABM; Hawkes; DL & Hurst exponents for absolute returns; Autocorrelation functions; Order flow imblance impact; Price Impact function; Spread w.r.t. time& A sensitivity analysis of the parameters with respect to the various stylized facts mentioned is performed to check the robustness of the parameters. \\
    
    \hline 
    \cite{wang2020} & DL  & PDFs of mid-price;  & Quantitative tests performed like KS test,Jarque-Bera test, Student t-test, t-statistic and p-values calculated for volatility measures (realized volatility, realized volatility per trade and intraday volatility)  \\
     \hline 
    \cite{shi2021limit, shi2021lob} & DL  & N/A  & Several baseline models like Ridge Regression, SVR, Random Forests, 1-layer feedforward Neural Networks etc. are used to test the performance of the model against other Machine Learning techniques. Further tests are conducted on the effectiveness of using an ODE-RNN by comparing the performance against LSTMs and GRUs. An ablation study is performed as well. \\
    \hline
    \cite{lim2021intra} & DL  & PDFs of mid-price returns  & Quantitative tests performed like KS test,Jarque-Bera test, Student t-test, t-statistic and p-values calculated for volatility measures (realized volatility, realized volatility per trade and intraday volatility)  \\
    \hline
    \cite{coletta2022learning} & DL  & PDF of log returns. PDFs of order type, time to fill, top of the book volumes, spread by time of day. Price Paths, Autocorrelation of returns and ACF of square returns.  & The simulation's unconditional and conditional PDFs of the stylized facts is compared with empirical PDFs and two baseline models.  \\
    \hline
   \highlightTwo{ \cite{prenzel2022dynamic}} & DL & PDFs (unconditional, conditioned by time of day, conditioned by market volatility) of order intensities by each order type, Price Paths & Qualitative tests performed for all stylized facts between real and simulated data.  \\
    \hline
    \highlightTwo{\cite{cont2023limit}} & DL &  PDF of Queue Size at best Ask/Bid; PDF of 3 levels of bid/ask queues; Average LOB Shape; Correlation of 3 levels of bid/ask queues; Price Paths; Price change probabilities &  Qualitative tests performed to test real vs simulated data for all stylized facts. Comparisons are also made between the GAN model the authors propose, a Poisson Process model and a Hawkes Process model. \\
    \hline
    \cite{nagy2023generative} & DL  & PDFs of returns, order type, arrival times & Perplexity scores are used to test the LLM. The simulation's unconditional and conditional PDFs of the stylized facts is compared with empirical PDFs.  \\
    
    \hline 
    \highlightTwo{\cite{ContLarrard2012}} & SPDE & Q-Q plot of inter-arrival times compared with exponential distribution, number of shares per event (to showcase clustering), Spread's (1 tick vs >1 tick) lifetimes, Joint PDF of bid and ask volumes, ACF of absolute order sizes, inverse of inter-arrival times, PDF of inter arrival times.  & Stylized facts are used to create priors on the model \\

    \hline
    \cite{korolev2015modeling}  & SPDE &  PDF of order arrival rate; ratio of intensities at bid/ask & Qualitative tests performed \\
    \hline
     \cite{chavez2017one} & SPDE &  Aymptotics of mid-price process; PDF of price; PDF of time for price change; Probability of price change conditional on current state &  N/A \\
     \hline
     \cite{gao2018hydrodynamic} & SPDE & Average shape of the order book, & Shapes of order book at various timestamps are compared qualitatively.  \\
    \hline
     \cite{rojas2020order}& SPDE & Joint distribution of best bid and ask size; PDF of spread; PDF of lifetime of spread ;Price Paths; Autocorrelation Function & Price Paths are compared qualitatively. \\
    \hline
    \cite{hambly2020limit} & SPDE &  Price Paths; Average LOB Shape &  Qualitative tests performed and volatility between empirical and simulated results is compared. Further this volatility is decomposed into two sources: exogenous movements and local imbalance.  \\
    \hline
    \highlightTwo{\cite{cont2021stochastic}} & SPDE &  Average LOB Shape &  Intraday Price Volatility are compared with empirical observations qualitatively   \\
    \hline

\end{tabularx}
\caption{Comparative Study}
\label{table:1}
\end{table}
\end{landscape}

\section*{Acknowledgements}

Konark Jain would like to acknowledge JP Morgan Chase \& Co. for his PhD scholarship. We are also grateful to Dr. Rama Cont and Dr. Stefan Zohren for reviewing and providing  comments  on  this  work. Finally,  we  are  grateful  to  the anonymous reviewers for their constructive feedback.

\section*{Disclosure Statement}

No potential conflict of interest was reported by the author(s).

\section*{Funding}

This work was supported by JP Morgan Chase \& Co.

\renewcommand*{\bibfont}{\small}
\printbibliography[
heading=bibintoc,
title={References}
]

\end{document}